\documentstyle[12pt,aasms4]{article}
\slugcomment{To be published in {\it The Astronomical Journal}}
\lefthead{Garilli et al.}
\righthead{BCM Surface Photometry}

\begin{document}

\title{TWO--COLOR SURFACE PHOTOMETRY OF BRIGHTEST CLUSTER MEMBERS}

\author{Bianca Garilli, Giorgio Sangalli, Stefano Andreon\altaffilmark{1}, Dario Maccagni}
\affil{Istituto di Fisica Cosmica del CNR, via Bassini 15, 20133 Milano, Italy\\
Electronic mail: bianca,sangalli,andreon,dario@ifctr.mi.cnr.it}

\and

\author{Luis Carrasco\altaffilmark{2} and Elsa Recillas\altaffilmark{2}}
\affil{Instituto Nacional de Astrofisica \'{O}ptica y Electr\'{o}nica, Apartado Postal 51 y 216, 72000 Puebla, Pue., Mexico\\
Electronic mail: carrasco,elsare@tonali.inaoep.mx}

\altaffiltext{2}{{\it Present address:} Osservatorio Astronomico di
Capodimonte, Napoli, Italy}
\altaffiltext{2}{Also Instituto de Astronomia, UNAM, Mexico D.F., Mexico}

\begin{abstract}
The Gunn $g$, $r$ and $i$ CCD images of a representative sample of 17 Brightest Cluster Galaxies (BCM) have been analyzed in order to derive surface
brightness and color profiles, together with geometrical parameters
like eccentricity and position angle. The sample includes both
X--ray and optically selected clusters, ranging in redshift from
$z=0.049$ to $z=0.191$. We find that BCMs are substantially well
described by de Vaucouleurs' law out to radii of $\sim60-80$ kpc,
and that color gradients are generally absent. Only in two cases
we find a surface brightness excess with respect to the $r^{1/4}$
law, which for A150 is coupled with a change in the $g-r$ color. 
The rest frame colors of BCMs
do not show any intrinsic dispersion. By parametrizing the
environment with the local galaxy number density, we find that
it is correlated with the BCM extension, i.e. BCMs with larger
effective radii are found in denser environments.    
\end{abstract}


\section {Introduction}
For many galaxies, the cluster environment is considered hostile,
often leading to interactions bound to
modify their structure and perturbing the quiet deployment of
their properties. This is at least what is happening {\it now},
what we observe in the dense cluster environments at the present
epoch in the form of perturbed morphologies. However, the cluster
environment has also had a great success in favoring the formation
of massive galaxies, specifically the most luminous galaxies
that are found. How has this occurred, and when? 
And what is the relation of these brightest cluster galaxies with the
surrounding cluster members, the cluster morphology, and the density of the
hot intracluster gas?\\

The study of brightest cluster members and of their structural
parameters has been traditionally pursued. The better
visibility of these objects has made them useful cosmological tools,
similar to supernovae in certain aspects. In order to use them as standard
candles, their properties have to be understood, and their
possible differences studied in terms of induced potential biases due to
similar yet somewhat different environments. This line of
research is still being followed, and would be further complemented
by an investigation of the properties of these giant galaxies
aimed at understanding their process of formation and evolutionary
status.\\

We have recently completed a photometric survey of the cores
of clusters of galaxies selected both through their X-ray
emission properties and optically through the density contrast
over the field (Garilli et al. \markcite{a1}1996, from now on Gar96). 
In that paper it was
shown that the same kind of passive evolution is shared by
the whole early type galaxy population in a cluster, and
no difference is evident between clusters selected in either way.
In this paper, we analyze the data set of Gar96 in an attempt to infer the
structural properties for a subsample of brightest cluster
members (BCM), and we try to demonstrate that there is a correlation
between the size of the BCM and the {\it local} galaxy density within the cluster,
as Andreon et al. \markcite{a2}
\markcite{a3}(1992, 1995) had suggested from a study of a much smaller sample.\\

This paper is organized in the following way: in
section 2, we describe the sample and the analysis tools; in
section 3, we present the properties of the studied galaxies, specifically their isophotal shapes,
their surface brightness profiles and color profiles;
in section 4, we discuss the inferred BCM properties in
relation to some other cluster observable parameters.\\   

Throughout this paper we adopt $H_0 = 50$ km~s$^{-1}$~Mpc$^{-1}$ and
$q_0=0.5$.

\section{The Data}
\subsection{The Sample}
The number of BCMs free from large superposed objects and for which our data allow a surface brightness
analysis out to a reasonable radius is 17, out of the 67 clusters
for which $g$, $r$ and $i$ photometry had been secured from the
2.1m telescope of the {\it Observatorio Astronomico Nacional} in
San Pedro Martir, Mexico.  Details of the observations are given
in Gar96. Table 1 lists the clusters in which
the BCMs are found, together with their redshifts, the Bautz--Morgan
type (Bautz \& Morgan 1970) and the richness class. Clusters marked ``blue`` or ``red``
are clusters in which the early type galaxy sequence is either
significantly bluer or redder than what expected from a properly
normalized Virgo color--magnitude relation (see Gar96).
These clusters could have followed an evolution slightly different 
from the one of the bulk of the clusters we studied.\\

The BCM subsample spans almost the full redshift range of the
cluster sample, $z$ ranging from 0.049 to 0.191, it includes
8 X--ray selected and 12 Abell clusters (Abell, Corwin \& Olowin 1989)
(3 clusters being
included both in the Abell catalog and in the EMSS (Gioia et al.
\markcite{a4}1990) catalog),
with B--M type I, I--II, II and II--III and richness
class from 0 to 2. However, if we can claim that our sample spans other a vast
range of cluster properties, we cannot say that we have an
unbiased sample. Our selection of
objects favors galaxies with large apparent radii. It is worth mentioning that the selection criteria of the main cluster sample
did not consider in any way the presence of a BCM of certain
characteristics, nor the observations were specifically tailored
to study their surface brightness properties.

\subsection{Data Analysis}
The 51 images containing the 17 BCMs were reduced and flux
calibrated as described in Gar96. Surface photometry, color
profiles and structural parameters have been obtained following
the procedures described in Andreon et al. \markcite{a3}(1995).\\

Briefly, the {\it sky background} has been determined in each image by
averaging the count rates in independent areas of $20\times 20$
pixels free from detectable objects and as distant as possible from
the BCM. We further required that the growth curves of each
BCMs actually reached saturation once assumed a sky background level. {\it Superposed or nearby objects} have been removed and
substituted with the interpolated value from nearby pixels prior
to any further analysis. In some cases the areas occupied by these
objects have been masked. It must be said than in any case these
removed objects are faint and small compared with the BCMs. It is worth
recalling that the absence of bright superposed objects was one of
the selection criteria for our subsample. {\it Eccentricities
and Position Angles} were computed by means of the STSDAS package
in IRAF. In order to keep the S/N ratio approximately constant
as the radius increases, we ran the procedures several times
over the same images rebinned by a proper factor. {\it Seeing
effects} were considered and the structural parameters have
been determined only by taking into account data points outside
an aperture 4 times the FWHM of the seeing disc. The average seeing
during our observations was $\sim 2\arcsec$. We have obtained the {\it eccentricity
and Position Angle} profiles outside the region affected by seeing and used
intensity weighted averages and center coordinates determined by STSDAS
to compute the {\it surface brightness profiles}. These profiles were
obtained by integration of the counts in elliptical coronae of
fixed eccentricity and position angle and of a step variable
according to S/N criteria. From the surface brightness profiles
in the three filters, {\it color profiles} have been extracted
and analyzed with respect to possible color gradients following
the method described by Sparks \& J\o rgensen \markcite{a5}(1993) and further
developed by Andreon et al. \markcite{a3}(1995).

\section{Results}
\subsection{Isophote Fitting, Eccentricity and Position Angle}
Ellipses are an excellent fit to the isophotes of every BCM 
studied here, as implied
by their $a_n$ Fourier coefficients that do not deviate
from 0 in any significant manner. However, only effects larger than 5\% could be detected in our data.\\

For all galaxies in our sample, the eccentricity and position angle profiles
do not significantly differ in the three filters. 
Figure 1 shows the eccentricity and position angle plots
as a function of the logarithm of the major axis expressed in kpc.
In Table 2,
for each BCM (column 1), we give the minimum and maximum
radius in kpc over which these parameters have been computed (column 2),    
an intensity weighted value of the eccentricity $\epsilon =
(1 - b/a)$ (column 3) and its rate of variation along the radius
expressed as $\Delta\epsilon$ kpc$^{-1}$ (column 4), an 
intensity weighted value of the position angle $\theta$ 
in degrees measured clockwise from the E--W direction (column 5)
and its rate of variation along the radius, also expressed as
$\Delta\theta$ kpc$^{-1}$ (column 6).\\

As one can notice, the eccentricities are never larger than 0.3, except in MS1558+3321 (A2145),
for which $\epsilon=0.5$. Only in the cases of MS0102+3255, A401 and MS1558+3321 (A2145) the eccentricity increases rapidly with the radius.
The isophotes are concentric within $0.6\arcsec$, which is the precision limit
that can be achieved with our data.
The rule is also to have a constant position angle of the isophotes, except in the
case of MS0013+1558 for which we detect a change.

\subsection{Surface Brightness Profiles}
In Figure 2 we present the BCM rest--frame surface brightness profiles in the $r$ bandpass
plotted as function of $r^{1/4}$. At small radii
our data are affected by seeing effects and consequently for those 
central regions the following arguments do not apply.
At large radii we
are limited by our flat--fielding and
S/N ratios. The arrows in the plots
indicate the galaxy rest frame surface brightness at a level 5 magnitudes fainter than the
sky background. In the case of A401 our
profile stops before this surface brightness limit as the edge of the CCD image was reached. The structural parameters have been derived considering
only those portions of the profiles, while, whenever meaningful, the profiles themselves
have been extended to fainter surface brightnesses in order to investigate the possible presence of extended halos.\\ 

For each galaxy in Figure 2 we also show as a continuous line the best fitting de Vaucouleurs' law (de Vaucouleurs 1948). In the fitting procedure, only
those data points that can be considered statistically independent
given the image seeing have been taken into account. For each
BCM, we report in Table 3 the results of the fits to the $r^{1/4}$
law. In column 2
and 3 we give the effective radius $r_e$ in kpc and the effective
surface brightness $\mu_e$ in  $r$ mag arcsec$^{-2}$, respectively.
The quoted errors represent the 90\% confidence levels given
by the fitting procedure. The surface brightness magnitudes were
corrected for k--dimming, galactic absorption and
atmospheric extinction following Schneider et al. \markcite{a6}(1983a) and
Stark et al. \markcite{a7}(1992), and also corrected for the $(1+z)^4$ cosmological
dimming.  Column 4 lists the obtained reduced $\chi^2$.  
Column 5 and 6 give the rest--frame mean $g-r$ and $r-i$
colors, and column 7 the absolute magnitude $M_r$ computed within
the 38 kpc radius. Generally, the de
Vaucouleurs' law  provides a satisfactory fit. We do not think that
our data justify any fit with a
different, more generalized law like the $r^{1/n}$ proposed by
Sersic \markcite{a16}(1968) and recently adopted by Caon et al.
\markcite{a18}(1993) and Graham et al.
\markcite{a17}(1996). 

From the radial light profiles presented in Figure 2, we can divide the sample BCMs into
three categories: (a) galaxies substantially well described by
de Vaucouleurs' law over the whole range of data points; (b)
galaxies showing what can be termed a tidal cut--off; and, (c)
galaxies showing a brightness excess over de Vaucouleurs' law.
We would place in category (a) (which we will not further discuss)
MS0002+1556, MS0013+1558, MS0102+3255, A180, A401, MS0301+1516, A671, A733,
and A1911. MS0037+2917(A77), A279, A399, MS1201+2824 and
MS1558+3321(A2145) fall into category (b), and finally A84,
A150, and MS0904+1651 (A744) are good candidates
for the presence of an extended halo (category (c)).\\ 

\subsubsection{Tidal cut--offs}
The profile of {\it MS0037+2917(A77)} mimicks a tidal cutoff beyond $\sim45$ kpc.
The apparent size of this galaxy is rather large with respect to
the CCD field of view and another bright S0 galaxy is present South
of the BCMs. Therefore, there is the possibility that the curvature in the profile (in any case, within 1 $\sigma$ of the de Vaucouleurs'
profile) might be due
to an overestimate of the sky background rather than to a real
physical effect.\\
In {\it A279}, a tidal cut--off could be
present outward of $\sim 60$ kpc at a very low significance level. Furthermore, this occurs beyond the
radius where systematic errors are important.\\
Similarly to A77, {\it A399} shows a surface brightness deficit
at large radii. Again, because of the large apparent size of
this galaxy with respect to the CCD size and of a removed object
along the galaxy major axis, we cannot confidently state that this deviation
from de Vaucouleurs' law is evidence of a tidal cutoff.\\
In the case of {\it MS1201+2824}, the impression of a surface brightness deficit
at large radii is probably due to the somewhat anomalous behavior of the
light 
profile with respect to the de Vaucouleurs' law.\\
As for {\it MS1558+3321 (A2145)},
the situation is similar to that of  A77 and A399, yet with even less
statistical significance.\\ 

\subsubsection{Extended halos}
Of the three BCMs showing some evidence of an extended halo, {\it A84} and
{\it MS0904+1651(A744)} show only scanty statistical significance
(the extrapolation of the best fit with de Vaucouleurs' law falls
within the 1 $\sigma$ error bars of the data points) even if for
A744 we were unable to trace the light 
profile in the rest--frame to a magnitude fainter
than $\mu_r\sim25$ mag arcsec$^{-2}$.\\
The case of {\it A150} is different, since this is also 
the most peculiar BCM in our sample. The light profile
shows a statistically significant excess outward of $\sim60$ kpc. Figure 3a
shows the residuals in the bi--dimensional image once a model
galaxy with parameters equal to the best fitting de Vaucouleurs'
is subtracted. As shown in the
figure, this excess does not correspond to that of
a classical halo, since it appears essentially
along the major axis.
Other peculiarity of this galaxy
is the presence of a double nucleus as can be seen in Figure 3b.\\   

The analysis of the surface brightness profiles of our BCM sample shows  that de Vaucouleurs' law provides an excellent description of
the surface brightness profiles
at radii between 10 and 45 kpc, where deviations are never
greater than 0.05 mag. At larger radii, deviations generally
remain of the order of the systematic errors, except in the case
of A150, which has a rather complex morphology.\\  

\subsection{Color Profiles}
$g-r$ and $r-i$ color profiles for all the BCMs have been
computed from the surface brightness profiles in the
three filters and are shown
in Figure 4. The method we employed to assess the presence
of color gradients takes into account the effect of incorrect
estimates of the sky background in one or both of the
images (see Andreon
et al. \markcite{a3}(1995) for a more detailed and complete explanation
of the method). The solid lines in Figure 4 represent the
expected color profiles in the absence of color gradients. Vertical
arrows indicate the position of superposed objects. 
Except in the case of A150, our plots are consistent with the
absence of color gradients.
This fact implies that color
gradients are generally not a characteristic feature of BCMs,
irrespective of the cluster selection criterion.\\

Let us now consider the $g-r$ radial profile of A150. The
delicate point of the adopted method is that a normalization procedure is
required. Whenever the data points scatter around the expected
no--color--gradient line, the concept of a normalization radius 
becomes 
irrelevant since no color gradient would be meaningfull
regardless of the radius chosen. 
In the case of A150, if we normalize the expected value for the $g-r$
color profile at a radius smaller than that at which we found 
excess light with respect
to the $r^{1/4}$ law, we would be led to state that the excess light 
is redder than that of the rest of the galaxy. On the other hand, if the
expected color profile is normalized at a radius where the
excess is present, the interpretation would be that the excess
and the central region of A150 share the same $g-r$ color
while at intermediate radii the galaxy is bluer. We have no
reasons to prefer one of the two interpretations of this color
profile; our data do tell us, however, that A150 is the only
BCM in the sample for which the $g-r$ color is not the same
everywhere.\\

\section{Discussion}
\subsection{Comparison with Other Authors}
Our sample includes seven galaxies which had been already studied
by other authors: A150 and A401 were studied by Malumuth \& Kirshner \markcite{a9} (1985)
in the $V$ filter, A671 is also in the Graham et al. \markcite{a17}
(1996) sample observed in the Kron--Cousins $R$ filter and MS0037+2917 (A77), A279, A399, A401, A671 and A733 are also
in the sample studied by Schneider et al. \markcite{a10}(1983b) and Hoessel
\& Schneider \markcite{a11}(1985) in the same Gunn $r$
filter.\\

Figure 5 compares our profiles in the $r$ band with those obtained
by Malumuth
\& Kirshner \markcite{a9}(1985) for A150 and A401. In the case
of A150, the agreement in surface brightness levels with Malumuth \& Kirshner's data is
excellent once the differences in filter
bandpasses are taken into account. However, our fitting procedure
yields a scale size $\sim 30\%$ smaller (and a
correspondingly brighter effective surface brightness), probably
because of the different range in radius over which the two fits are carried out. In the case of A401 we determined an effective
radius $\sim 25\%$ larger than Malumuth \& Kirshner. Porter et al. \markcite{a15}(1991) already
compared the surface brightness profile they obtained for A401
with Malumuth \& Kirshner's and showed that they derived a steeper
profile and a higher surface brightness for the central regions.
Our profile has a slope similar to that of Malumuth \& Kirshner and 
also a value of the
central surface brightness similar to that of Porter et al.'s.
We should mention that the profile we derived for A671 is in very good agreement with
Graham et al.'s \markcite{a17}(1996) as well as are the parameters
obtained with the $r^{1/4}$ law.\\

For the other commonly studied BCMs, a comparison with Hoessel \& Schneider's \markcite{a11}(1985) results shows
differences  in the effective radii of the order of $7\arcsec$ to $10\arcsec$.
Several causes
could explain these differences: the way seeing effects
are taken into account, the outermost radius to which the fit
is applied,
the way eccentricity is taken into account.
However, the relation between effective radius
and surface brightness of BCMs found by Hoessel \& Schneider \markcite{a11}(1985) 
is also a very good fit to the whole of our data. We conclude
that a detailed, object--to--object comparison is rather meaningless because it depends on the data quality, and on the reduction and
analysis techniques adopted. However, the statistical properties
of the samples are much more robust and reproducible regardless of
the authors, therefore these are really representative of the samples.

\subsection{Intrinsic Colors}
In Table 3, columns 5 and 6 give the rest--frame $g-r$ and $r-i$
colors, and column 7 the absolute magnitude $M_r$, all computed within
a radius $r=\sqrt{ab}=38$ kpc.\\

The average rest--frame $r-i$ color of the BCMs in our sample
is $<r-i> = 0.28$ with a sample standard deviation of 0.05. This value
agrees with the one given by
Schneider et al. \markcite{a6}(1983a) and the dispersion around the mean
is of the same order as the photometric accuracy obtained for
this color (Gar96). In this case the mean and the median 
values do coincide.\\ 

The average rest--frame
$g-r$ color is $<g-r> = 0.31$ with a sample standard deviation of 0.09. Once the shift in the $g$ filter of our photometric system with
respect to that of Schneider et al. \markcite{a6}(1983a) (Gar96) is taken into
account,
also this value is compatible with the BCM rest--frame colors
of those authors. The standard deviation is however about a factor of 2
higher than expected on the basis of the photometric errors. If
we use the k--corrections given by Frei \& Gunn \markcite{a20}(1994)
for E galaxies,
all galaxies have slightly redder rest--frame colors with an average $<g-r>=0.35$ and a sample standard deviation of 0.07. Although the standard
deviation remains higher than expected, it seems difficult to
claim that BCMs show an intrinsic dispersion in this color. We
recall that also in Andreon et al.'s \markcite{a3}(1995) BCM sample, no
intrinsic dispersion in the $B-V$ color was found.

\subsection{Relation with the Environment}
Andreon et al. \markcite{a3}(1995) showed that a correlation
existed between the BCM effective radius and the cluster richness.
In that paper, Abell's richness class was used as a measure of
the environment galaxy density. For the present sample, we
cannot use Abell richness class as this is not available
for the X--ray selected clusters. However, we can correlate BCM
effective radii with the local galaxy density. The cluster complete
catalogs of Gar96 allow us to count galaxies brighter than
$M_r=-18$ in circles of projected radii equal to 200 kpc at the
cluster redshift and centered on the BCM itself. Plotted
against the effective radius (Figure 6), this measure of
the local galaxy density confirms and quantifies the
result of Andreon et al. \markcite{a3}(1995). Shallower
BCMs are found in environments, which, at the present epoch, are
locally denser.\\

\subsection{BCMs as standard candles: the Tolman (1930) test}
Sandage \& Perelmuter (1990) suggest that BCMs can be used to test 
whether the Universe is expanding or not. To check the cosmological
dimming, they looked for a
redshift dependence of the galaxy surface brightness at a fixed metric radius.\\ 

For the galaxies studied by Sandage \& Perelmuter \markcite{a22}(1990), the observed scatter in the 
relation between $\log(1+z)$ and $\mu$, i.e. the scatter in
the cosmological dimming, once the Malmquist bias (Malmquist 1920)
is removed (see Sandage \& Perelmuter 1990 for details), is $\sim 0.25$ mag arcsec$^{-2}$. For our BCM sample, in the
range $0.05<z<0.2$, we find a corresponding scatter of 0.22 
mag arcsec$^{-2}$. 
The slope of the linear regression between $\log(1+z)$ and $\mu$, before
Malmquist corrections, is 3 times larger than the Tolman signal for 
Sandage \& Perelmuter's (1990)
samples, yet for our sample is of the same amplitude as the Tolman signal.
This suggests that our sample is only slightly affected by the Malmquist
bias. This fact is probably due to the different sample compositions, 
and, most probably, to the fact that in our sample about half of the
BCMs are there because they are members of clusters serendipitously detected
in an X--ray survey. Although the cluster X-ray luminosity is correlated
with the BCM absolute magnitude (Valentijn \& Bijleveld \markcite{a23}1983),
their surface brightness at a fixed
metric radius is only weakly dependent on it, implying that through X--ray selection the
amount of bias introduced on the quantity to test as redshift
increases is quite small. We suggest
that the Tolman test could give significant results if carried
out on a proper sized sample of BCMs drawn from purely X-ray selected clusters. Unfortunately, the number of BCMs in X--ray selected clusters
in our sample is still too small and extends to relatively low redshift
values
for the Tolman test to be successfull.

\section{Conclusions}
The multicolor surface brightness photometry of this mixed sample
of 17 BCMs with redshifts extending to $z\sim0.2$ shows that:

\begin{itemize}
\item The surface brightness profile 
of these BCMs is
generally well described by de Vaucouleurs' law out to radii of
$\sim60-80$ kpc. Structural peculiarities
such as changes in isophotal position angle and/or ellipticity are
largely absent in our sample.
\item Although normal and small size ellipticals do show
color gradients (Sparks \& J\o rgensen \markcite{a5}(1993)), these gradients
are a rather rare feature in BCMs. The only color gradient
detected in our sample corresponds to A150. This galaxy also shows 
a brightness excess with respect to the $r^{1/4}$ law along
the major axis and a double nucleus. Andreon et al. \markcite{a3}
(1995), who analyzed color profiles in the same way as we did, 
found that 2 BCMs in a sample of 9 show 
gradients in the $V-i$ color (and 2 out of 8 BCMs in the $B-V$ color). 
We could thus infer
that $\sim10\%$ of BCMs presents measurable color gradients in
broad--bandpasses redwards of the 4000 \AA~break. 
\item BCM $g-r$ and $r-i$ colors (integrated within 38 kpc radius)
do not show any
intrinsic dispersion. If an intrinsic dispersion is present in the
bluer color, this is of the same order as the uncertainties in the
k--corrections.
\end{itemize}

This {\it normality} of BCMs which tends to assimilate them to the
more general class of elliptical galaxies does not remove their
main peculiarity: the large scale size of many of them (about half
of the BCMs we studied have effective radii $r_e>40$ kpc). Galaxies
so large and bright ($M_r<-24$) are found only in clusters
of galaxies. Furthermore, we have shown that their linear dimensions
scale with the local galaxy density as determined at the
present epoch. The sizes do not seem to be related either to the
cluster morphology, or to the way clusters are selected.
Bautz--Morgan type I clusters possess BCMs among the less extended (A180, with an effective radius of only 22 kpc)
and among the most extended ones in our sample (A401, $r_e=115$ kpc).
It is true that only one of the five X--ray selected clusters (i.e., 
clusters which have been missed in the Abell catalog) has a BCM
with an effective radius larger than 40 kpc (MS1201+2824, $r_e=48$
kpc), but this is naturally explained by a selection effect: these
clusters are not found in the Abell catalog because they are not
very rich, and therefore their BCMs are not very extended. Nor is
the X--ray luminosity of the clusters correlated with the size of
the BCM galaxy.\\

Merging is considered the most likely mechanism for the formation
of these giant galaxies. It also provides a way to dilute any
color gradients (White \markcite{a21}1980) (which we do not
generally observe). Two different cD formation scenarios have
been investigated: either cDs are the product of repeated mergers after cluster virialization (Malumuth 1992; Bode et al. 1994),
or the merging events leading to the cD formation occurred
before or during cluster formation (Merritt 1988). There is evidence of merging events occuring in clusters at the present epoch,
as shown by the peculiar E galaxy IC1182 in the Hercules cluster
which is the result of a recent interaction between an elliptical
and a spiral galaxy (Tarenghi \& Maccagni, private communication).
Generally, simulations
consider the luminosity of the merger products and, if this is
a few times $L^*$, the resulting galaxy is identified as a cD, i.e.
the most luminous galaxy at the center of the cluster. Dynamical
arguments lead us to consider the cluster as a whole, that is
to say, the entire cluster galaxy population comes into play, 
mainly since it is largely concentrated within the core radius. Furthermore,
it is sometimes assumed that the large extent of the cD galaxy is
mainly due to the presence of a halo. The relation we found
is between {\it de Vaucouleurs effective radius and
the galaxy number density within a 200 kpc radius from the
BCM itself}. It is therefore a relation between a quantity
which measures the slope of the surface brightness profile,
independently of the presence of an extended halo (which does
not seem to be a common feature in our sample),
and a quantity which measures the density of the environment
as it is now. Both quantities are presumably the end points of
evolutionary processes, which might not necessarily occur with the same time scales. The implications of Malumuth's (1992) simulations for cDs, provided the dynamical friction in the real
universe is as efficient as in his code, are, among others, that
cDs are a relatively recent phenomenon or that they are not formed in their present environment. Our data would
tend to exclude this last possibility, and, because of the
generally dull appearance, do probably suggest that some time
has elapsed since the last dramatic event.
This seems to be in contrast with the high peculiar
velocities sometimes shown by cDs (Hill \markcite{a41} et al. 1988, Sharples, Ellis \& Gray 
\markcite{a42} 1988, Zabludoff, Huchra \& Geller \markcite{a43} 1990, 
Malumuth \markcite{a44} et al. 1992). However, 
substructure is a common feature of many clusters (Geller \& Beers 
\markcite{a45} 1982, 
Jones \& Forman \markcite{a46} 1984, Bird \markcite{a47} 1994), and when it is taken into account in the dynamical analysis of the clusters,
the peculiar velocities of the central galaxies tend to
disappear (Bird 1994). Our data thus seem to suggest that the growth in size of the BCMs is governed by the {\it local} environment and that,
in many cases, the dominant galaxy and the concentration of galaxies surrounding it represent, at least for a non negligible amount of time, a dynamical entity with respect to the whole cluster. 
It would be
extremely interesting to probe the dynamical status of this
sample with respect to the very local and to the overall cluster
environment since this would be a powerful diagnostic tool for
the formation scenarios.

\acknowledgements
The authors warmly acknowledge the support of the OAN staff both in San 
Pedro Martir and Ensenada. The Time Allocation Committee of the
UNAM--OAN is thanked for the generous allocation of telescope time.
S.A. acknowledges the award of a CNR fellowship. We are grateful
to the referee, E.M. Malumuth, for his useful suggestions which
have helped us to improve the contents and presentation of the paper.

\newpage
\begin{deluxetable}{lcccr}

\tablewidth{0pc}
\small
\tablecaption{Cluster Characteristics}
\tablehead{
\colhead{Cluster}      & \colhead{redshift} &
\colhead{Bautz--Morgan}      & \colhead{Richness}  &
\colhead{Notes}\\
\colhead{id}           & \colhead{$z$}   &
\colhead{Type} & \colhead{Class} &
\colhead{}}

\startdata
MS0002+1556 & 0.116 & \nodata &
\nodata & \nodata \nl
MS0013+1558 & 0.083 & \nodata &
\nodata & \nodata \nl
MS0037+2917 (A77) & 0.072 & I &
1 & \nodata \nl
A84 & 0.103 & II &
1 & \nodata \nl
MS0102+3255 & 0.080 & \nodata &
\nodata & \nodata \nl
A150 & 0.060 & I--II &
1 & \nodata \nl
A180 & 0.135 & I &
0 & blue \nl
A279 & 0.080 & I--II &
1 & \nodata \nl
A399 & 0.072 & I--II &
1 & \nodata \nl
A401 & 0.075 & I &
2 & \nodata \nl
MS0301+1516 & 0.083 & \nodata &
\nodata & red \nl
A671 & 0.049 & II--III &
0 & \nodata \nl
A733 & 0.116 & I &
1 & \nodata \nl
MS0904+1651 (A744) & 0.073 & II &
1 & blue \nl
MS1201+2824 & 0.167 & \nodata &
\nodata & \nodata \nl
A1911 & 0.191 & II--III &
2 & \nodata \nl
MS1558+3321 (A2145) & 0.088 & \nodata &
0 & \nodata \nl
\enddata

\end{deluxetable}

\newpage

\begin{deluxetable}{lcccrc}
\tablewidth{0pc}
\small
\tablecaption{BCM Eccentricities and Position Angles}
\tablehead{
\colhead{Cluster}      & \colhead{$a_{min}-a_{max}$} &
\colhead{$\epsilon$}      & \colhead{$\Delta\epsilon$ kpc$^{-1}$}  &
\colhead{$\theta$}      & \colhead{$\Delta\theta$ kpc$^{-1}$}\\
\colhead{id}           & \colhead{kpc}   &
\colhead{} & \colhead{} &
\colhead{degrees} & \colhead{}}

\startdata
MS0002+1556 & 17--65 & 0.30 & 0.003 &
65 & 0.6 \nl
MS0013+1558 & 10--58 & 0.20 & 0.006 &
--38 & 1.2 \nl
MS0037+2917 (A77) & 9--49 & 0.20 & 0.008 &
--26 & \nodata \nl
A84 & 13--41 & 0.18 & \nodata &
165 & \nodata \nl
MS0102+3255 & 10--47 & 0.23 & 0.012 &
28 & \nodata \nl
A150 & 9--49 & 0.18 & 0.008 &
105 & \nodata \nl
A180 & 16--44 & 0.22 & 0.005 &
103 & \nodata \nl
A279 & 12--30 & 0.13 & 0.007 &
46 & \nodata \nl
A399 & 7--51 & 0.27 & 0.006 &
128 & \nodata \nl
A401 & 8--55 & 0.29 & 0.010 &
59 & \nodata \nl
MS0301+1516 & 12--37 & 0.26 & 0.003 &
--14 & \nodata \nl
A671 & 4--36 & 0.19 & 0.005 &
157 & \nodata \nl
A733 & 6--41 & 0.11 & \nodata &
265 & 0.8 \nl
MS0904+1651 (A744) & 9--43 & 0.06 & \nodata &
185 & \nodata \nl
MS1201+2824 & 22--44 & 0.30 & \nodata &
95 & 0.4 \nl
A1911 & 22--41 & 0.30 & 0.001 &
76 & 0.3 \nl
MS1558+3321 (A2145) & 11--61 & 0.50 & 0.013 &
82 & \nodata \nl
\enddata

\end{deluxetable}

\newpage

\begin{deluxetable}{lrccccc}
\tablewidth{0pc}
\small
\tablecaption{BCM Parameters}
\tablehead{
\colhead{Cluster}      & \colhead{$r_e$} &
\colhead{$\mu_e$}      & \colhead{$\chi^2_{red}$} & 
\colhead{$g-r$}        & \colhead{$r-i$}        & 
\colhead{$M_r$} \\
\colhead{id}           & \colhead{kpc}   &
\colhead{$r$ mag arcsec$^{-2}$} & \colhead{} & 
\colhead{}             & \colhead{}             & 
\colhead{}}

\startdata
MS0002+1556 & $36.7^{+3.0}_{-1.1}$ & $23.43^{+0.15}_{-0.10}$ &
0.17 & 0.35 & 0.29 & --23.65 \nl
MS0013+1558 & $24.6^{+1.0}_{-0.5}$ & $22.71^{+0.05}_{-0.05}$ &
1.70\tablenotemark{a} & 0.39 & 0.28 & --23.72 \nl
MS0037+2917 (A77) & $35.9^{+2.5}_{-3.1}$ & $23.41^{+0.20}_{-0.15}$ &
1.04 & 0.37 & 0.27 & --23.58 \nl
A84 & $58.8^{+5.2}_{-2.9}$ & $23.97^{+0.20}_{-0.05}$ &
0.16 & 0.34 & 0.29 & --23.83 \nl
MS0102+3255 & $18.1^{+1.7}_{-1.4}$ & $22.59^{+0.50}_{-0.50}$ &
0.53 & 0.30 & 0.28 & --23.28 \nl
A150 & $39.3^{+3.1}_{-1.4}$ & $23.47^{+0.15}_{-0.12}$ &
7.58\tablenotemark{a} & 0.26 & 0.33 & --23.69 \nl
A180 & $21.8^{+1.0}_{-1.9}$ & $22.56^{+0.07}_{-0.19}$ &
0.12 & 0.29 & 0.30 & --23.59 \nl
A279 & $44.5^{+2.2}_{-0.8}$ & $23.45^{+0.10}_{-0.05}$ &
0.16 & 0.20 & 0.35 & --23.90 \nl
A399 & $67.8^{+2.5}_{-3.0}$ & $23.97^{+0.20}_{-0.10} $ &
0.67 & 0.20 & 0.27 & --23.94 \nl
A401 & $115.0^{+7.6}_{-4.8}$ & $24.39^{+0.25}_{-0.07}$ &
0.04 & 0.42 & 0.24 & --24.25 \nl
MS0301+1516 & $21.1^{+1.1}_{-1.0}$ & $23.09^{+0.06}_{-0.05}$ &
0.29 & 0.40 & 0.30 & --22.99 \nl
A671 & $42.4^{+3.1}_{-2.2}$ & $23.29^{+0.35}_{-0.20}$ &
3.25\tablenotemark{a} & 0.33 & 0.13 & --24.06 \nl
A733 & $28.8^{+1.3}_{-2.1}$ & $22.31^{+0.20}_{-0.35}$ &
0.32 & 0.21 & 0.18 & --24.36 \nl
MS0904+1651 (A744) & $30.2^{+3.1}_{-2.2}$ & $22.91^{+0.08}_{-0.10}$ &
0.81 & 0.21 & 0.29 & --23.71 \nl
MS1201+2824 & $48.4^{+2.1}_{-1.8}$ & $23.28^{+0.10}_{-0.05}$ &
1.04 & 0.25 & 0.24 & --24.10 \nl
A1911 & $66.6^{+5.9}_{-2.3}$ & $23.66^{+0.10}_{-0.15}$ &
0.48 & 0.29 & 0.23 & --24.38 \nl
MS1558+3321 (A2145) & $75.5^{+9.4}_{-7.8}$ & $24.56^{+0.40}_{-0.06}$ &
0.50 & 0.31 & 0.31 & --23.62 \nl
\enddata

\tablenotetext{a}{see text}
\end{deluxetable}

\newpage

\figcaption[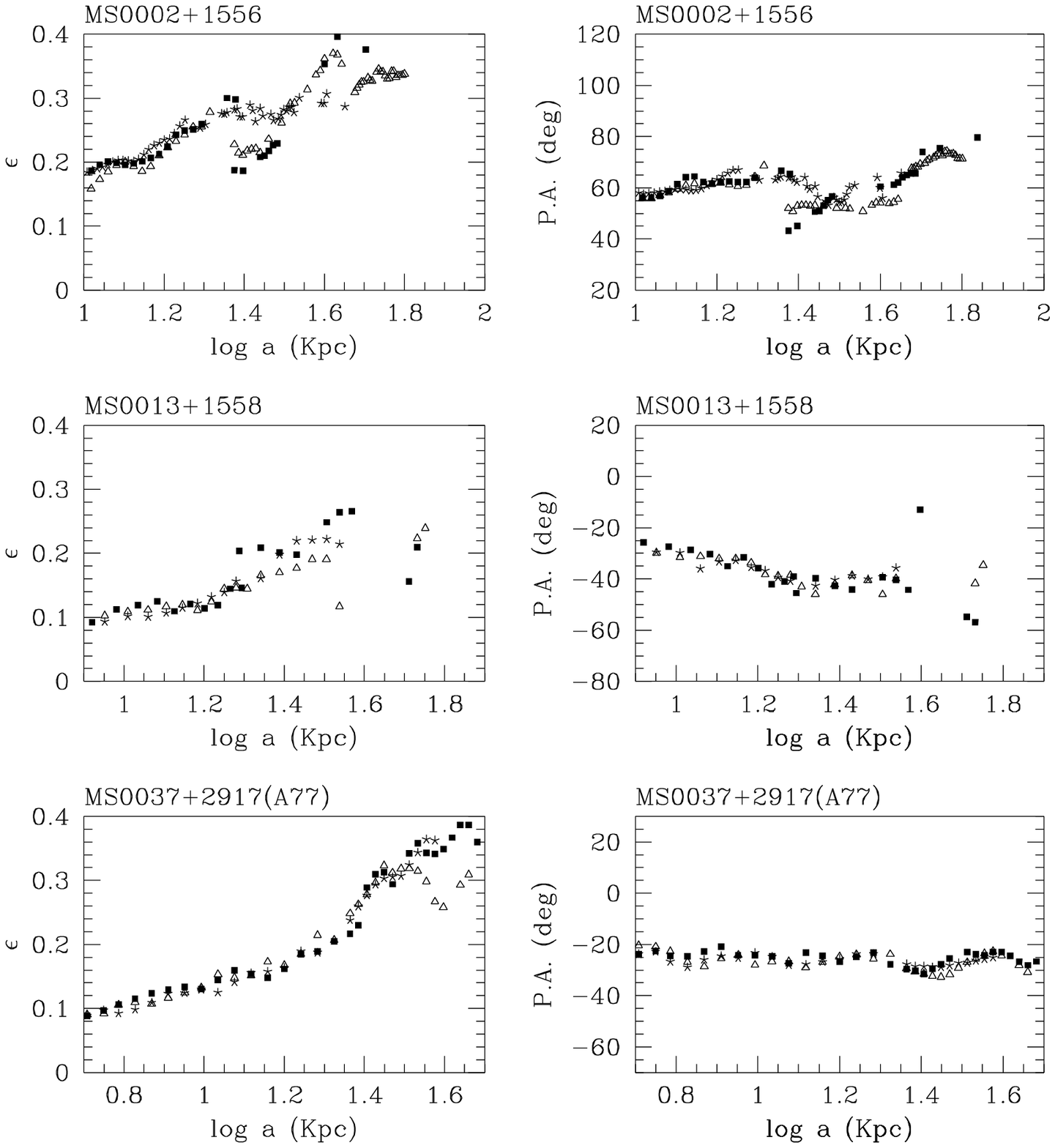]
{BCM eccentricity $\epsilon=(1-b/a)$ and position angle P.A. as a function of the
logarithm of the semi--major axis $a$ expressed in kpc. {\it Open triangles}: $g$ band;
{\it full squares}: $r$ band; {\it stars}: $i$ band.}

\figcaption[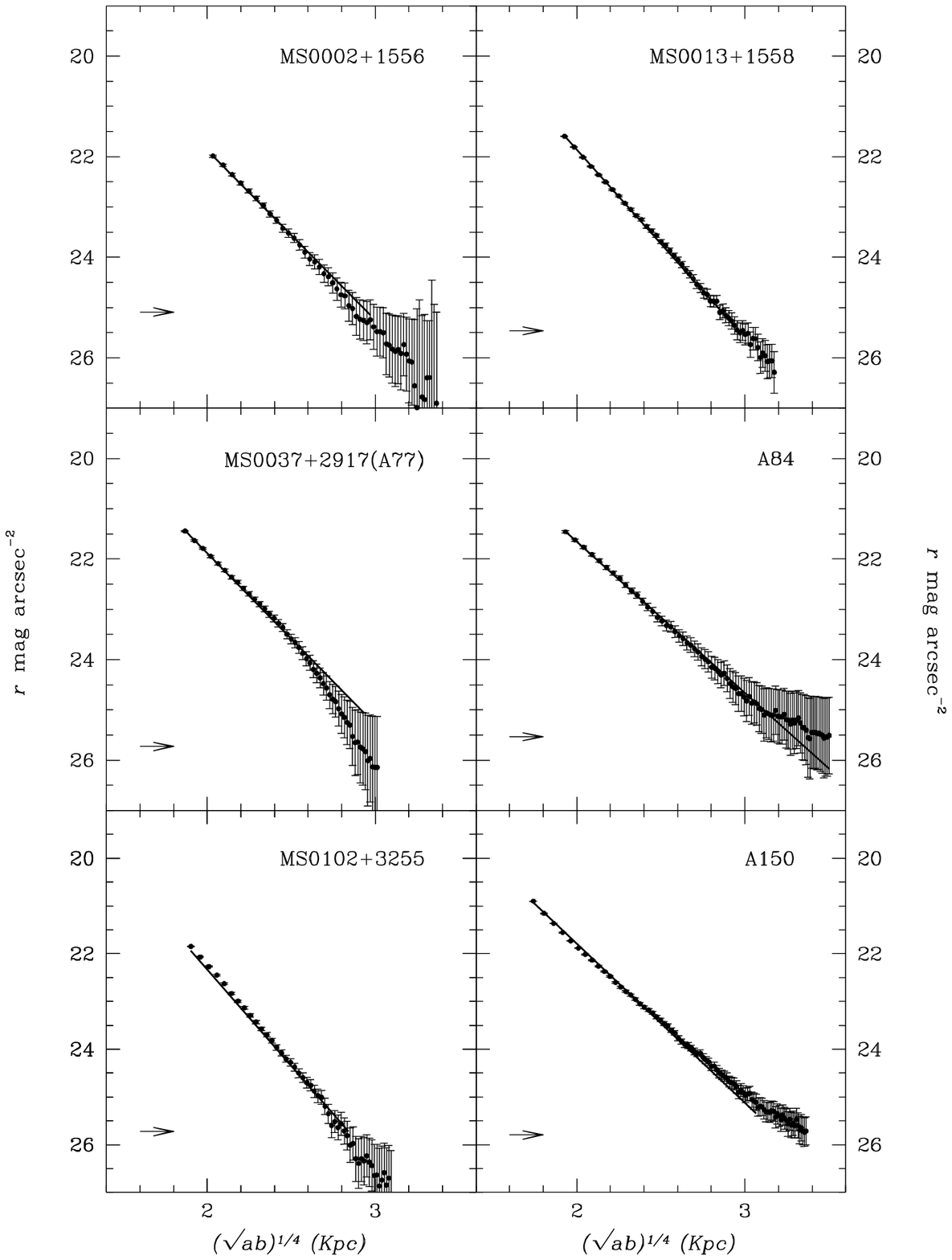]
{BCM rest frame surface brightness profiles in the $r$ filter
plotted as a function of $r^{1/4}$, where $r=(ab)^{1/2}$. The
continuous line is the best fitting de Vaucouleurs' law. The fit
has been performed on every other data point beyond 2 times the FWHM of
the seeing disc so that they can be considered independent, and within
the radius where the observed galaxy surface brightness is 5
magnitudes fainter than the sky background (indicated by a horizontal 
arrow). Error bars are the composition of the counting 
statistics and of the sky background r.m.s. error. The profiles have been plotted outward as far as large and bright objects
set in or the edge of the field of view is reached.}

\figcaption[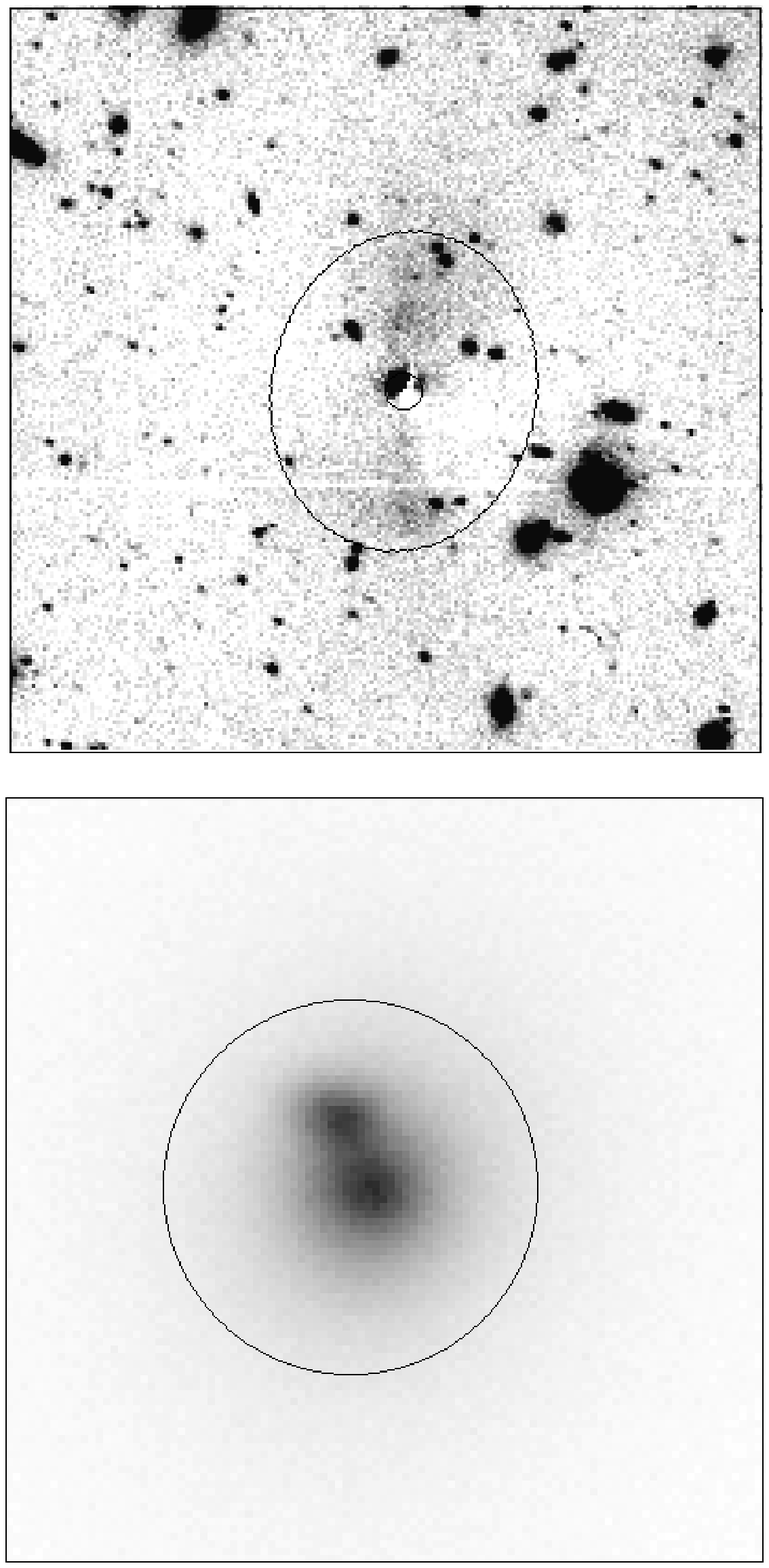]
{$(a)$ The residual image of A150 once a model galaxy
with the best fitting de Vaucouleurs' parameters has been
subtracted. The ellipse is the isophote at $(ab)^{1/2}=2r_e$
and the inner circle is two times the seeing disc. The excess surface brightness along the major
axis is evident. $(b)$ The A150 $r$ image shows the presence of a
resolved double nucleus. Again, the circle is two times the seeing disc.}

\figcaption[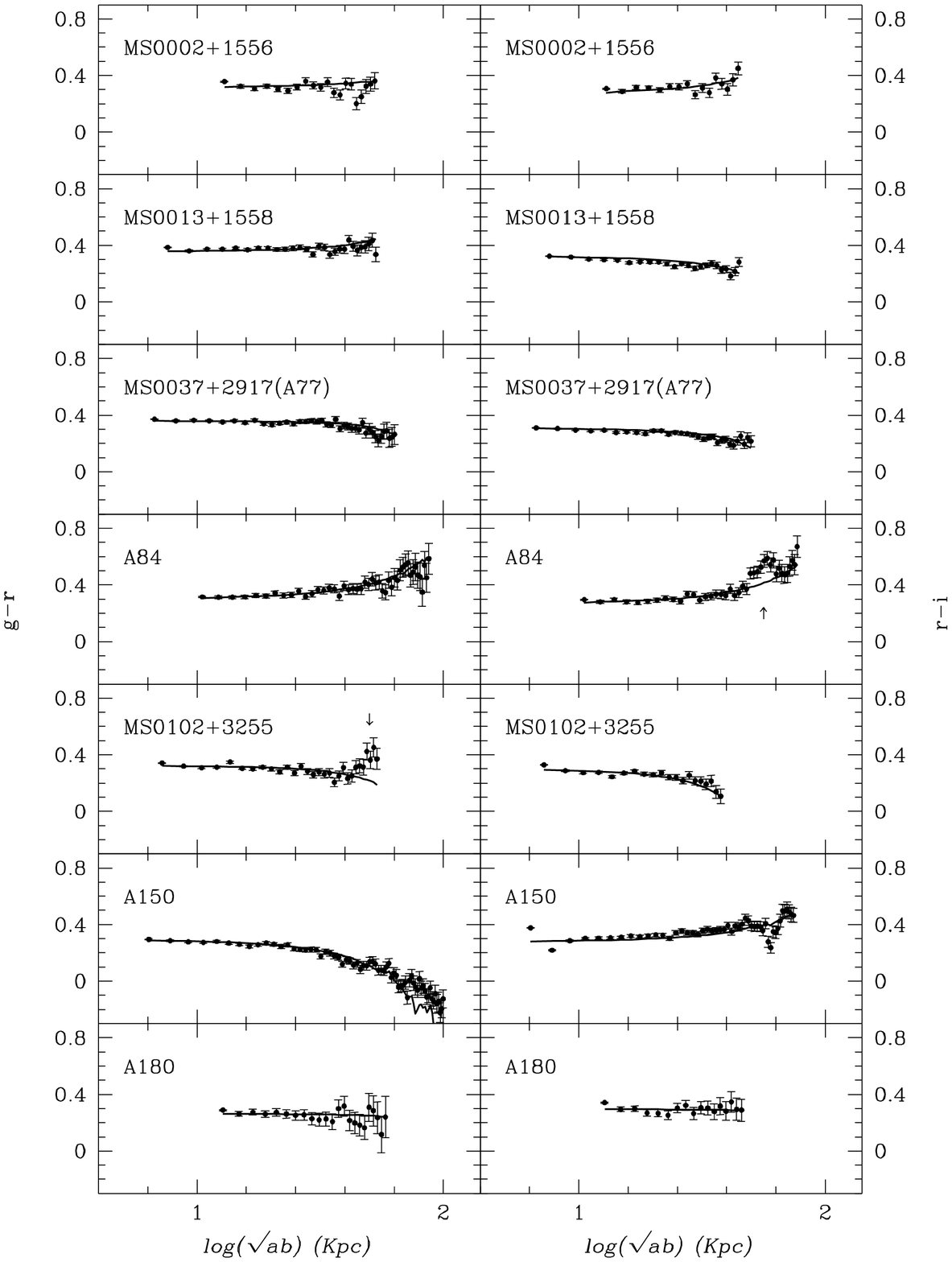]
{$g-r$ (left) and $r-i$ (right) color profiles of
the sample BCMs. The continuous line is the expected profile
in the absence of color gradients (see text and Andreon et al.
(1995) for a complete explanation of the method). Vertical arrows 
indicate the position of superposed rather bright objects.}         

\figcaption[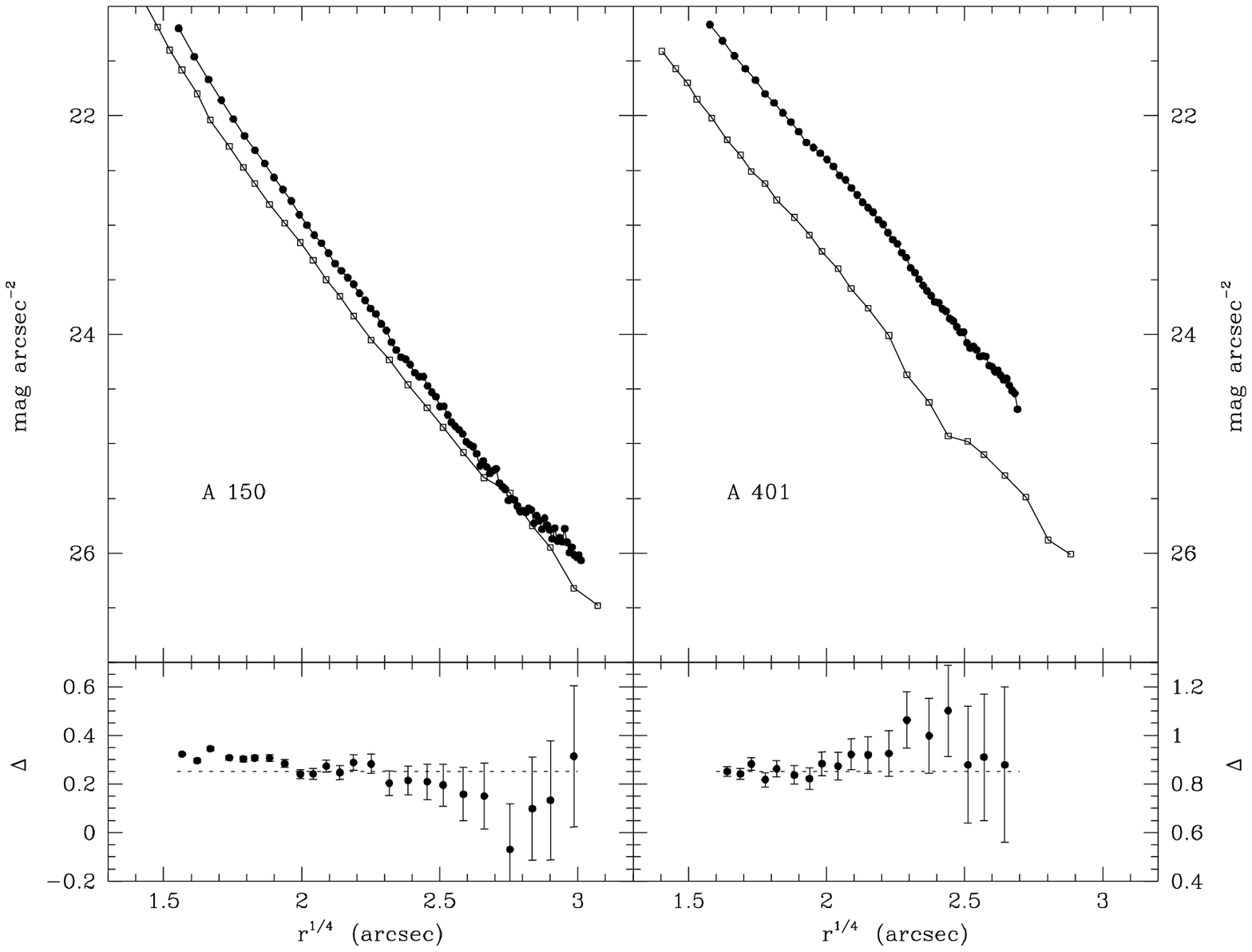]
{Comparison of our $r$ band surface brightness profiles of A150
and A401 (upper panels, full dots) with the $V$ band profiles
of Malumuth \& Kirshner (1985) (upper panels, open squares). The
lower panels show the difference between the profiles we obtained
and those obtained by Malumuth \& Kirshner. Our data points (and errors)
have been resampled at the same radii as Malumuth \& Kirshner's data.}

\figcaption[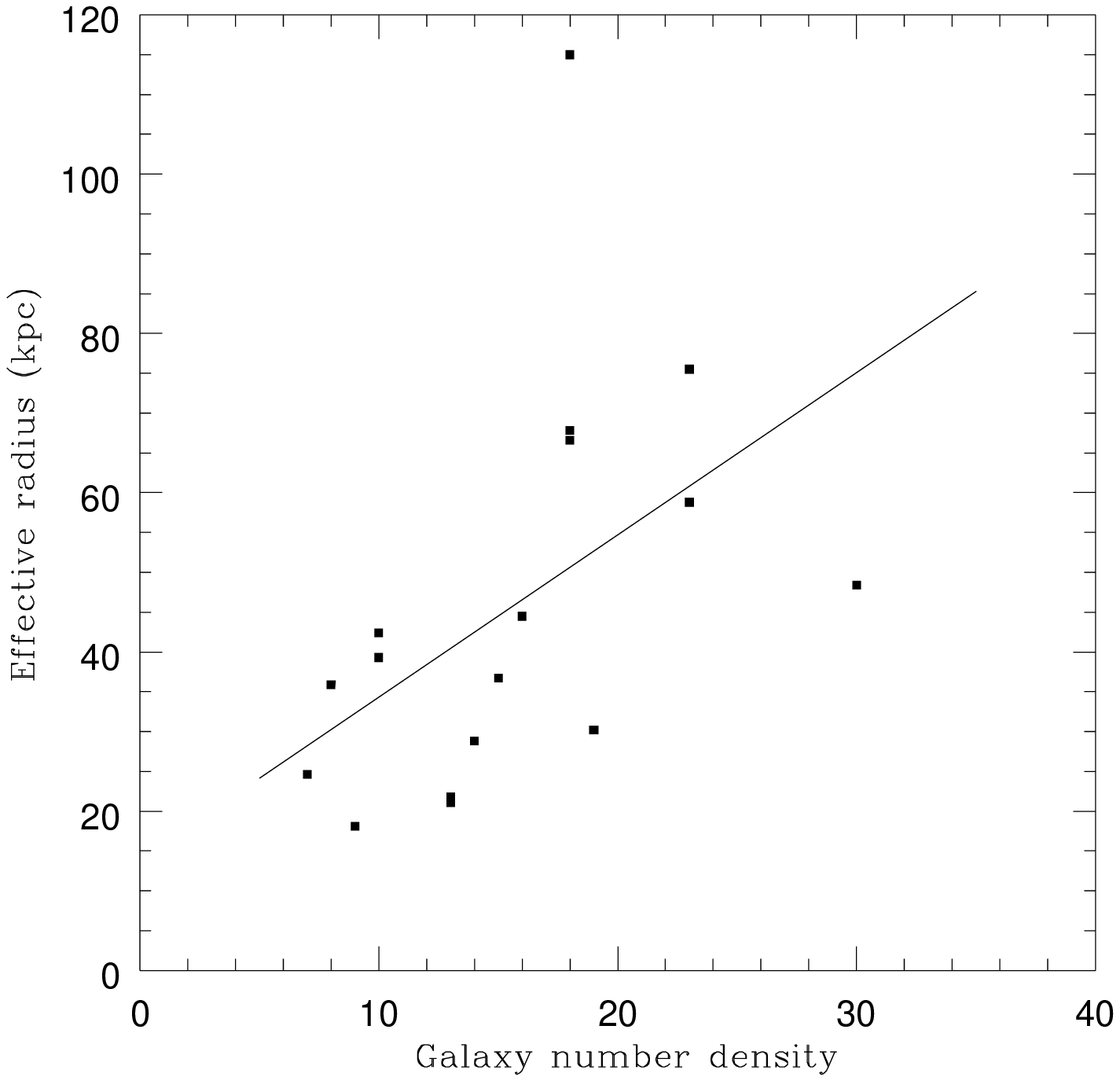]
{Effective radius $r_e$ $vs.$ the number of galaxies
brighter than $M_r=-18$ within a 200 kpc radius from the center
of the BCM. The continuous line is the least square fit through the
data points, yielding a correlation coefficient $r=0.50$.}

\end{document}